\documentclass[aps,pre,twocolumn,superscriptaddress,,amsmath,amssymb,]{revtex4-1} 
\usepackage{graphicx}
\usepackage{xcolor}
\usepackage{dcolumn}
\usepackage{bm}

\usepackage[utf8]{inputenc}
\usepackage[T1]{fontenc}
\usepackage{mathptmx}
\usepackage{etoolbox}
\usepackage{soul}

\begin{document}

\title{ Evolution of Shielding Cloud  Under Oscillatory External Forcing  in Strongly Coupled Ultracold Neutral Plasma}

\author{Mamta Yadav}
\email {ymamta358@gmail.com} 
\affiliation{Department of Physics, Indian Institute of Technology Delhi, Hauz Khas, New Delhi 110016, India}
		
\author{Aman Singh Katariya}
\affiliation{Department of Physics, Indian Institute of Technology Delhi, Hauz Khas, New Delhi 110016, India}

\author{Animesh Sharma}
\affiliation{Department of Physics, Indian Institute of Technology Delhi, Hauz Khas, New Delhi 110016, India}

\author{Amita Das}
\email {amita@iitd.ac.in, amitadas3@yahoo.com}
\affiliation{Department of Physics, Indian Institute of Technology Delhi, Hauz Khas, New Delhi 110016, India}

\begin{abstract}

This paper investigates the dynamics of crystalline clusters observed in Molecular Dynamics (MD) studies conducted earlier  [Yadav, M., et al. Physical Review E, 107(5), 055214(2023)] for ultra-cold neutral plasmas. An external oscillatory forcing is applied for this purpose and the evolution is tracked with the help of MD simulations using the open source LAMMPS software. Interesting observations relating to cluster dynamics are presented. The formation of a pentagonal arrangement of particles is also reported.

\end{abstract}
\maketitle

 \section{ Introduction}
Ultracold plasmas have emerged as an exciting frontier research field for experimental \cite{killian1999creation,kulin2000plasma}, numerical simulation \cite{kuzmin2002numerical}, and theoretical  \cite{glinsky1991guiding} investigation. It offers a unique environment for studying complex atomic and molecular systems. These low-temperature plasmas have drawn significant attention due to their potential applications in fields such as precision spectroscopy, quantum computing, and astrophysics. In this work, we carry out the Molecular Dynamics simulation using an open-source code LAMMPS software to understand the behavior of crystalline shielding cloud around a static uniform highly charged particle ( observed earlier \cite{yadav2023structure}), under the influence of an external spatially homogenous and oscillatory electric field. We also seek an understanding of a dynamic shielding response from the strongly coupled plasma medium, when the static particle has oscillatory charge of the form $q(t) = q_0+q_1 sin(\omega_0 t)$ on it. Such a disturbance in the medium can also be looked upon as the insertion of a biased Langmuir probe in experiments, with the biasing voltage being time-dependent.

In the work, the electrons and ions are chosen to interact through the long-range Coulomb potential along with an additional short-range repulsive LJ potential. The LJ potential essentially prevents the recombination of electrons and ions, thereby containing the blowing up of simulation by restricting the particles from undergoing close encounters. 

 The strong and weak coupling regimes are characterized based on the value of a dimensionless parameter known as the coulomb coupling parameter ($\Gamma$) defined as $\Gamma = Q^2/K_b T a$. Here $Q$ is the charge on the particle, $K_b$ is the Boltzmann constant, $T$ is the temperature of the species, and $a$ is the inter-particle separation. If $\Gamma$ is greater than one, the system is in a strongly coupled regime; otherwise, it is in a weakly coupled regime. Thus, the strongly coupled regime can be achieved by (i) increasing the charge $Q$, (ii) decreasing the temperature $T$,  and (iii) reducing the inter-particle distance $a$.  The strong coupling condition, therefore,  requires stringent considerations of very dense plasmas (encountered in the context of laser-ionized solid targets)  or at very low mili kelvin level temperatures which have been achieved through laser cooling in the context of ultra-cold plasmas.

It is also important to note that a strong coupling regime can also be achieved if one can have particles in a medium having a very high charge. This has indeed been achieved in the context of dusty plasmas where the dust particles immersed in the plasma medium, acquire very high electronic charge through electron attachment.  The strongly coupled behavior of dust particles gets manifested even at room temperature and normal densities.  The particle dynamics in such cases can also be visually tracked.  Thus dusty plasmas have proven to be excellent systems for studying the structures and dynamics in the strongly coupled regime. 
The field has evoked a strong interest and has led to a variety of experimental and simulation work. For instance, crystal and cluster formation, their rotation, vortex motion, instabilities, oscillations, and transport have been some of the prominent themes lately \cite{cheung2003rotation,barkan1995laboratory,praburam1996experimental,merlino1998laboratory,gerasimov1998formation,hwang1998consequences,nunomura1998confinement,nunomura1999instability, tiwari2012kelvin, das2014suppression,dharodi2022kelvin, veeresha2005rayleigh, wang2000self,samarian2001self}. The linear waves and nonlinear structures supported by the strongly coupled dusty plasma medium \cite{kumar2019coupling,kumar2018spiral,das2014exact}, nonlinear structures \cite{kumar2017observation,das2014collective,tiwari2015molecular}, phase transitions \cite{maity2019molecular}, and interplay of collective and single particle dynamics \cite{maity2018interplay} have been explored. Various studies also show the presence of chaos in the plasma medium \cite{fan1992observations,sheridan2005chaos,sheridan2010transition}. In \cite{sheridan2010transition}, T. Sheridan et al. have shown experimentally the transition to chaos as a function of the amplitude of a periodic driving force for two different driving frequencies, with three particles in a driven dusty plasma.  The dynamics of small dust clusters and the presence of chaos in them have been shown in \cite{maity2020dynamical,deshwal2022chaotic}. Thus an enormous literature exists in the context of dusty plasmas. 

 Studies in the neutral strongly coupled plasma have also attracted considerable interest. It is a complex domain for research requiring sophisticated experimental arrangements and diagnostics as one has to deal with superdense matter and/or ultracold regimes. Such kind of matter can also be naturally found in super dense stars \cite{van1991dense}. Significant advancements have also been made in the creation of ultracold plasma experimentally by using the technique of laser cooling. Ultracold neutral plasmas (UNP) have been generated in various atomic and molecular systems, such as strontium (Sr) \cite{nagel2003magnetic,killian2005absorption}, calcium (Ca) \cite{cummings2005fluorescence}, rubidium (Rb) \cite{feldbaum2002coulomb,wilson2013density}, cesium (Cs) \cite{robinson2000spontaneous}, and even in a molecular system (nitric oxide, NO) \cite{morrison2008evolution}.  For ultracold plasmas, the typical electron and ion temperature is of the order of millikelvin to microkelvin.  A wide range of characteristics and properties of ultracold plasma have been extensively investigated, including collective modes \cite{fletcher2006observation,castro2010ion}, expansion dynamics \cite{cummings2005fluorescence,kulin2000plasma,robicheaux2002simulation,pohl2004kinetic,laha2007experimental}, collisions, both elastic and inelastic \cite{bannasch2012velocity,wilson2013density,killian2001formation}, correlations \cite{mcquillen2013imaging,pohl2004coulomb,simien2004using,mazevet2002evolution}, formation of Rydberg atoms \cite{killian2001formation}, thermodynamics and transport properties \cite{tiwari2017thermodynamic}, etc.

 Earlier studies \cite{yadav2023structure} have demonstrated the formation of classical bound structures containing both kinds of particles. We have also observed the formation of crystalline patterns in the process of shielding an externally applied potential.  The 
 addition of an external static high-charged particle in simulations essentially mimics a biased probe.  Our aim here is to examine the dynamical behavior of the crystalline shielding patterns in the strongly coupled ultracold plasma regime when they are perturbed by external oscillating fields. It is demonstrated that the dynamics of the shielding process are extremely sensitive to the magnitude of the applied field. A detailed analysis showed that rather than chaos the dynamics exhibit stochastic and random in this particular context. In addition, we also carry out studies on the dynamics of the shielding crystalline pattern when the biased potential is time-dependent and oscillatory. Again this is achieved by inserting an oscillatory charged ($q(t) = q_0+q_1 \sin(\omega_0 t)$) particle in the medium.  Interestingly, the choice of this time-dependent charge leads to the formation of a pentagonal shielding cloud in some cases rather than the hexagonal form that is usually observed.

 The paper has been organized as follows. In Sec. II, the simulation detail has been discussed. In Sec. III, we study the dynamics of crystalline shielding clusters in the presence of an external uniform oscillatory electric field ($\vec{E}=E_0 sin(\omega_0 t) \hat{x}$). The clusters rotate and change their direction of rotation at random intervals. A time series analysis is carried out to show that the process is random and stochastic. A parametric dependence on ($E_0$ and $\omega_0$) on the dynamics has been presented. In Sec. IV, we study the behavior of crystalline shielding cloud in the presence of an oscillatory biased probe mimicked by a spatially stationary external high-charged article inserted in the medium. The charge is chosen to vary sinusoidally.   In section V, we briefly summarize and conclude the key findings of our studies.   

  \label{intro}

\section{ MD Simulation Details}
\label{mdsim}

In this study, two-dimensional (2D)  molecular dynamics (MD) simulations have been carried out using an open-source code LAMMPS \cite{plimpton1995fast}. We have used the parameters associated with the ultra-cold electron-ion plasma regime. In the simulation box, each particle interacts with the  other particle through a long-range pair Coulomb potential ($V_{pC}$) and a short-range repulsive core Lennard-Jones (LJ) potential ($V_{LJ}$) as given by

\begin{equation}
V(r_{lm}) = V_{pC}+ V_{LJ}
\label{total_potential}
\end{equation}
where,
\begin{equation}
{V}_{pC}= \frac{Q_{m}}{4\pi\epsilon_{0}r_{lm}}
\label{Coulomb_potential}
\end{equation}

\begin{equation}
{V}_{LJ}= 4\epsilon\left[\left(\frac{\sigma}{r_{lm}}\right)^{12}-\left(\frac{\sigma}{r_{lm}}\right)^{6}\right]
\label{lj_potential}
\end{equation} 
 This short-range LJ potential is used to avoid the recombination of opposite charge species. Here, $Q_m$ represents the charge on a typical  $m^{th}$ particle. Also, $r_{lm}$ defines the distance between $l^{th}$ and $m^{th}$ particles. Here, $\epsilon$ and $\sigma$ are the LJ potential parameters where $\epsilon$ describe the depth of the LJ potential well and $\sigma$ is the distance at which particle-particle LJ potential is zero. Other forms of potentials are also used in the literature to resolve the blowing-up of short-range Coulomb attraction \cite{tiwari2017thermodynamic,deutsch1977nodal,hansen1978ground}. For instance, the potential   $({-e^2}/{r})(1-exp(-e^2/(A a)^2))$ \cite{tiwari2017thermodynamic} also avoids the recombination of opposite kind of charge species. Here, the electron-ion repulsion length scale is set by the adjustable parameter $A$. For this form of potential and the LJ potential, the minima in the potential profile occur at some finite value of interparticle separation ($r$ ). The presence of minima in the potential leads to the formation of bound states. Some other forms of potential \cite{deutsch1977nodal,hansen1978ground} (i.e. $({-e^2}/{r})(1-exp(-Cr))$) are also used in the context of dense plasma. Here $C$ represents the inverse De-Broglie wavelength. It should be noted that in this case, there is no minimum in the potential profile at any finite value of $r$. \\ 
 
\begin{figure}[hbt!]
 \includegraphics[scale=0.32]{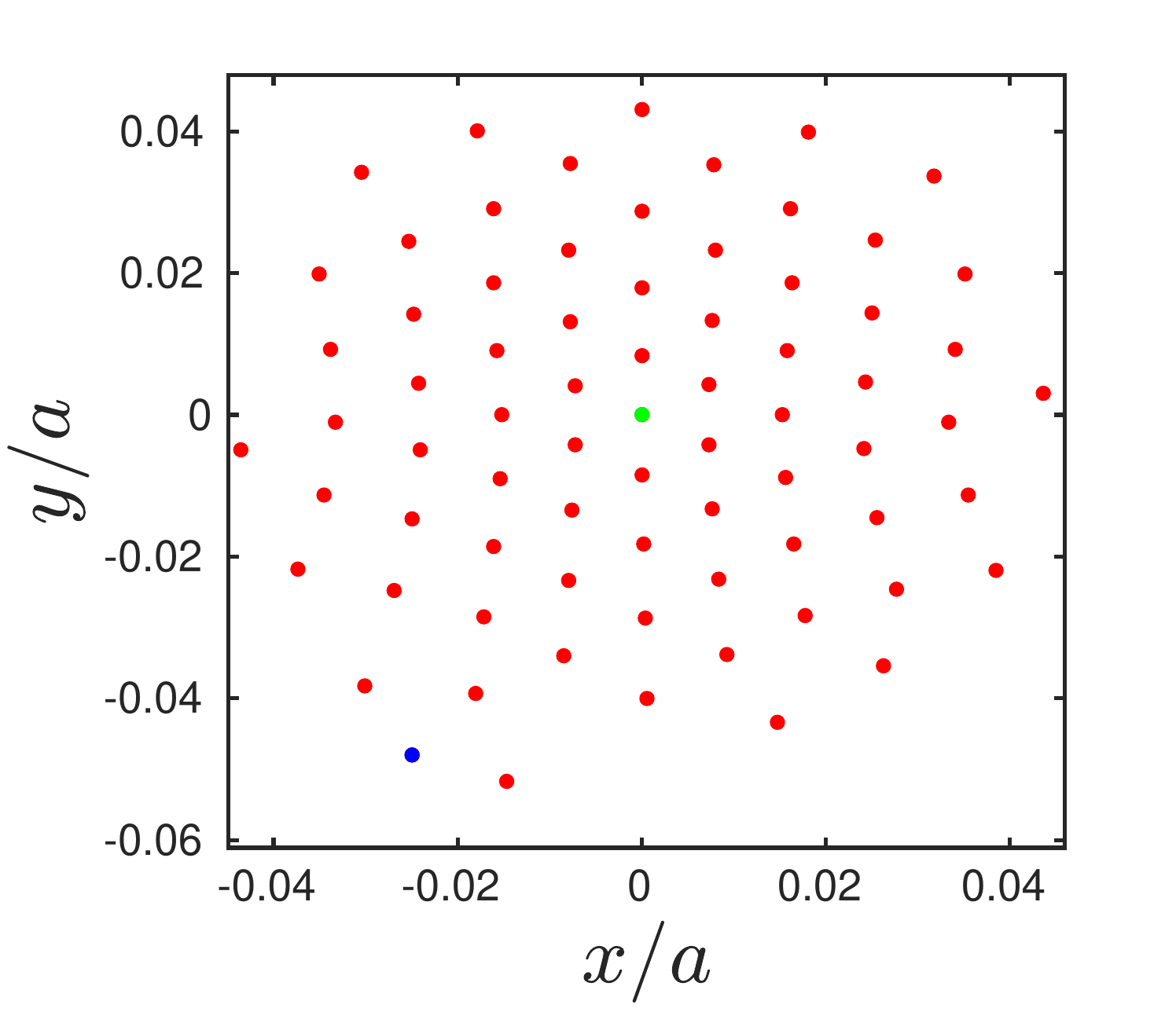}
     \caption{(a) The crystalline structure formed by electrons near the external perturbation. The red, blue, and green color represents the electrons, ion, and external perturbation.  }
  \label{Fig:initial}
 \end{figure}

 The equation of motion for any $l^{th}$ particle is described as
  
\begin{equation}
 m_l\frac{d^2\mathbf{r}_l}{dt^2}= -Q_l \nabla \sum_{m=1}^{N-1} V_{pC} - \nabla \sum_{m=1}^{N-1} V_{LJ},
\label{equation of motion}
\end{equation}
where $N$ is the total number of particles in the simulation box.
We have carried out a two-dimensional (2D) study and have chosen the length of the simulation box as $L_x = L_y = 1.4\times10^{-3} $ meters in the $\hat x$ and $\hat y$ directions, respectively. In our previous work \cite{yadav2023structure}, we carried out the study of the shielding process exhibited by a strongly coupled medium. For this purpose, a static high charge was inserted in the medium. Oppositely charged particles were observed to accumulate around the static charge,  at specific locations forming a crystalline hexagonal structure.  This has been illustrated in  Fig.\ref{Fig:initial} for ready reference as its evolution will be studied subject to external time-dependent perturbations here. 
The red and blue dots correspond to electrons and ions respectively and the green dot is the static test charge inserted in the medium. The charge and mass of the green inserted particle are chosen to be $1.6\times10^{-17}$ C and $10^4$ times the mass of electrons, respectively. The charge ($Q_i$) and mass ($M_i$) of the ions are taken to be $1.6\times10^{-19}$ C and  $100$ times the mass of electrons, respectively. This is essentially to reduce the simulation time and computational cost. The geometry being two dimensional the areal density $n$ and the average inter-particle distance $a$ are related by  $a = 1/\sqrt{n\pi}$. The areal density is chosen to be $1.0\times10^{10} m^{-2}$ and the corresponding average inter-particle distance is, therefore,  $5.64\times10^{-6}$ m. The periodic boundary conditions are imposed in both $\hat x$ and $\hat y$ directions. For our system, we consider the temperature of electron and ion species to be the same i.e. $T_e=T_i=0.1$ $K$. The choice of these parameters leads to the value of the  Coulomb coupling parameter $\Gamma$ of $22.8$. The value of $\Gamma$ is greater than unity for both the electron and ion species and hence both of them are in the strongly coupled regime. The typical normalizations that are used to display our plots depict length and time scales by the average inter-particle distance ($a$) and the inverse of electron-plasma frequency ($\omega_{pe}^{-1}$), respectively. To track the dynamics of lightest species (i.e., electrons) we have chosen the time step $8\times10^{-8}\omega_{pe}^{-1}$. 

 The cutoff distance for Coulomb pair potential interaction amongst particles in terms of average inter-particle distance is chosen to be $r_c=20a$. The parameter $\sigma$ and the normalized parameter $\epsilon$ that is associated with LJ potential is chosen to be $10^{-2}a$ and $558 K_BT$, respectively. The Nose-Hoover thermostat (NVT) \cite{nose1984unified,hoover1985canonical} has been used to achieve the required thermal equilibrium that is associated with the desired value of Coulomb coupling parameter $\Gamma$.

\section{ Behaviour of shielding cloud in the presence of external oscillatory and spatially homogeneous forcing}
\begin{figure}[hbt!]
 \includegraphics[height = 8.0cm,width = 9.0cm]{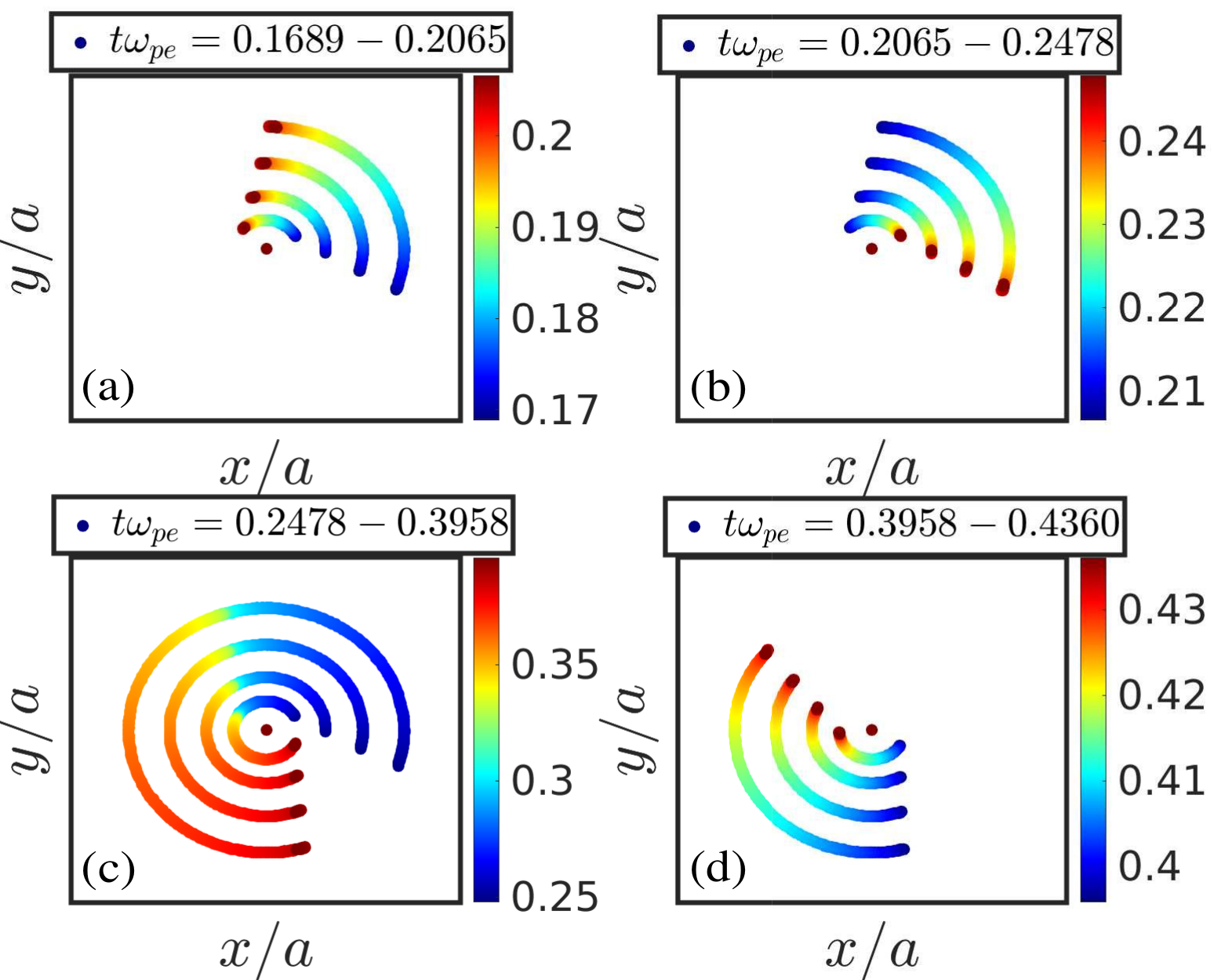}
     \caption{ Trajectory of one particle from inner four shells over the time duration $t\omega_{pe} = 0.1689-0.4360$. Different colors from blue to red show the time evolution of particles. Here, the amplitude and frequency of the applied field are $100$ V/m and $10\omega_{pe}$, respectively  }
  \label{Fig:trajectory}
 \end{figure}

We first investigate the behavior of the shielding crystalline cloud depicted in Fig.\ref{Fig:initial} in the presence of an externally applied oscillatory electric field 
 given by the expression $\vec{E} = \vec{E}_0 \sin(\omega_0 t) $. This field is chosen to be directed along  $\hat x$. Here  $E_0$ and $\omega_0$ are the amplitude and frequency of the applied oscillatory electric field, respectively. We observe that the particles arranged in the crystalline lattice execute a circular oscillatory motion around the static charge placed at the center under the influence of this additional forcing field. The rotational motion changes from clockwise to anti-clockwise and vice versa after a certain interval which seems quite random. For instance, when the amplitude and the frequency of the oscillatory electric field as $100$ V/m (which translates to $5.64\times 10^{-4}$ V over a distance of interparticle lattice spacing in which particles are arranged )  and $10\omega_{pe}$, respectively; the intervals at which the rotational motion switches depicted in Fig.\ref{Fig:trajectory} is fairly random.  Fig.\ref{Fig:trajectory} shows the trajectory of four different particles chosen from four different shells around the central particle. Each of the subplots in this figure has been plotted for a duration for which the particles rotate along the same direction. Thus,  subplot (a) refers to a duration from 
 $t \omega_{pe} = 0.1689 $ to $0.2065$ (the color blue to red in each subplot shows the time evolution ) in which the rotation is anti-clockwise. Thereafter, the trajectory is plotted in subplot(b) for the duration $\omega_{pe} t = 0.2065 $ to $0.2478$ where the rotation is clockwise. The subplots(c) and (d) are also for the duration when the rotation changes. 
 It is clear that the oscillatory rotation has no connection with the frequency of the applied force, nor does it depend on the natural plasma frequency. 

\begin{figure}[hbt!]
 \includegraphics[height = 8.5cm,width = 9cm]{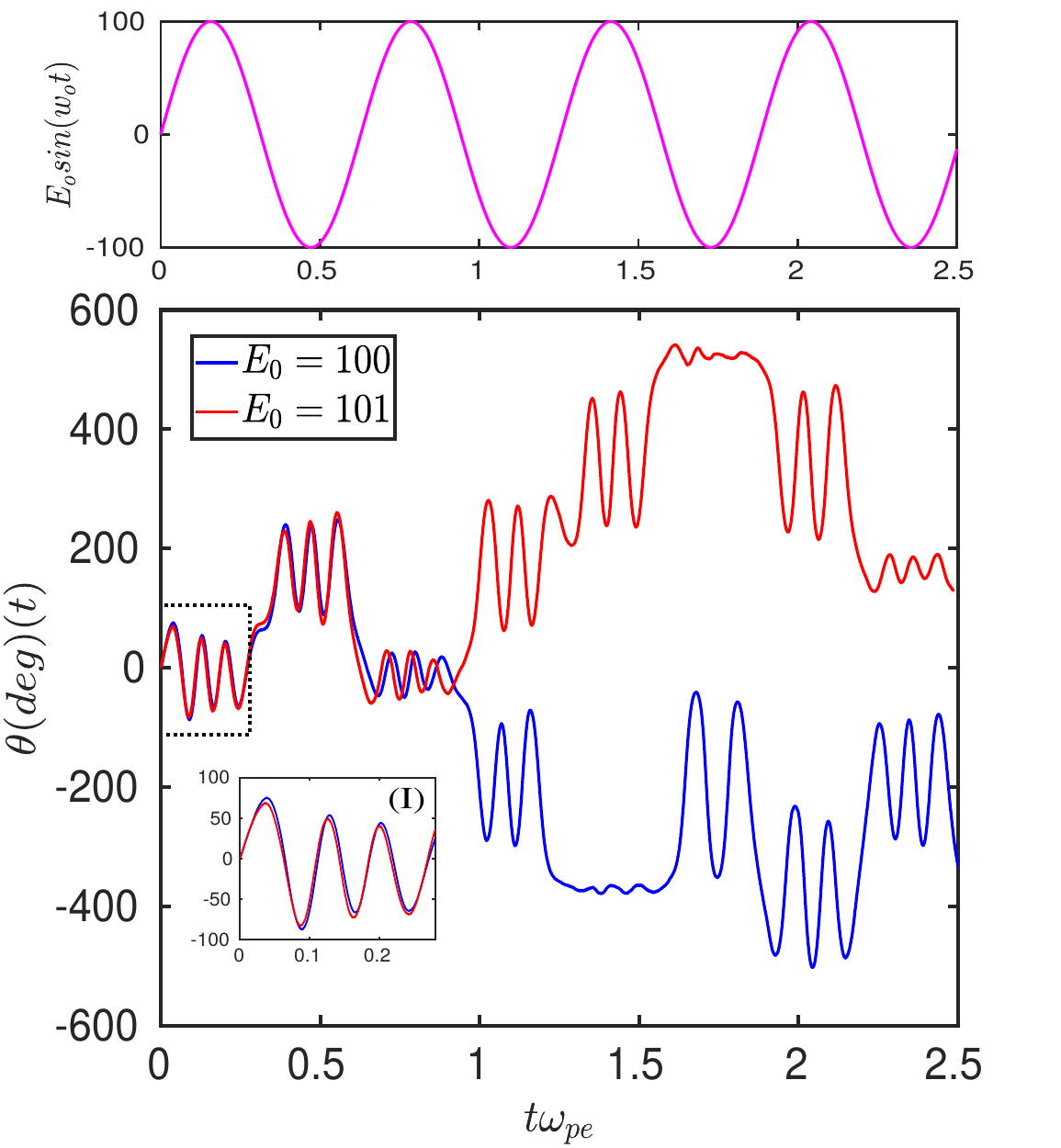}
     \caption{The time evolution of angular displacement for one of the particles from the cluster with changing the amplitude of oscillatory electric field $E_0$.The blue and red curve corresponds to amplitude $100$ V/m and $101$ V/m, respectively. The magenta curve shows the variation of the oscillatory electric field ($E_0 sin(\omega_0 t)$) with the normalized time ($t\omega_{pe}$). }
  \label{Fig:theta}
 \end{figure}

 We have also plotted the angular displacement $\theta(t)$  as a function of time for a single particle picked up randomly from the cluster,  in Fig.\ref{Fig:theta}. 
 The angular displacement $\theta$ is calculated as
\begin{equation}
    \theta(t) = \left[tan^{-1}\left(\frac{Y_i(t)}{X_i(t)}\right)-tan^{-1}\left(\frac{Y_i(t_o)}{X_i(t_o)}\right)\right]
    \label{equation:theta_eqn}
\end{equation}
where $t_o$ is the arbitrarily chosen initial time. whereas, $X_i$ and $Y_i$ are described as $X_i =x_i-L/2$ and $Y_i=y_i-L/2$, respectively.
 The two colors in the plot represent cases for slightly different forcing amplitudes (e.g. blue for $E_0 = 100$ and red for  $E_0=101$).  For both cases, the applied frequency was chosen to be the same. Even such a small difference in forcing alters the trajectories very significantly. The dynamics are very sensitive to the choice of parameters in the forcing field. 
  
\begin{figure}[hbt!]
   \includegraphics[scale=0.4]{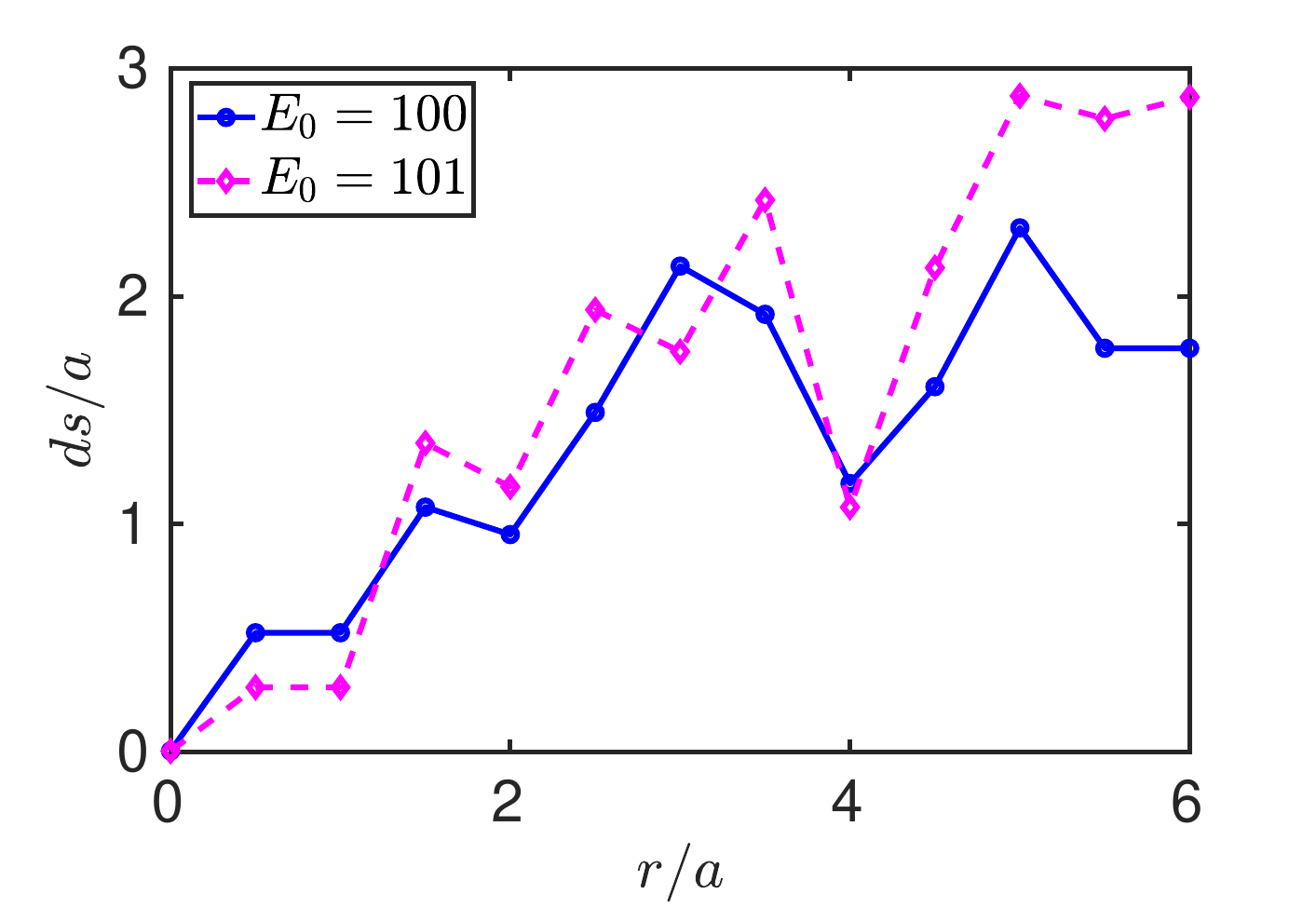}
     \caption{Variation of average displacement $ds$ with the radius for two different amplitude $100$ V/m (solid blue curve) and $101$ V/m (dashed magenta curve) of oscillatory electric field. }
  \label{Fig:displacement}
 \end{figure}

However, there is also regularity in the sense that the whole cluster primarily exhibits a rigid oscillation. This is evident from the movie "cluster.avi" \cite{movie1} attached as supplementary material. It also shows up in the plot of average displacement ($ds$) of particles positioned at various radius $r$ during a specific time interval $\tau$. The average displacement $ds$ is calculated as 
 \begin{equation}
     ds(r) =  \frac{r}{N_p} \sum_{j=1}^{N_p} [\theta_j(t+\tau)-\theta_j(t)]
 \end{equation}
 Here $N_p$ is the number of particles present within a shell between radius  $r$ and  $r+dr$ from the center of the cluster. For a perfect rigid displacement,  $ds$ should linearly increase with $r$ ($ds = r d\theta$). We can observe from Fig.\ref{Fig:displacement}, that the trend of both the curves corresponding to different amplitudes of forcing function different amplitudes $100$ V/m (solid blue curve) and $101$ V/m (dashed magenta curve) is linearly increasing with $r$. Thus on average, the cluster exhibits a rigid oscillation with some disturbances. 
It is evident thus that though the order in the spatial crystalline arrangement of particles remains preserved the rotational dynamics exhibit significant randomness. We explore it further in the following subsection through a time series analysis of the velocity of the particles under the action of this force. 

\subsection{Time series analysis} 
 The angular velocity  ($V_\theta$) of any particle in the cluster in terms of the $V_x$ and $V_y$, the $x$ and $y$ components of the velocity is given by 
 \begin{equation}
     V_\theta = -V_x sin\theta + V_y cos\theta
 \end{equation}

 \begin{figure*}
\centering
   \includegraphics[scale=0.4]{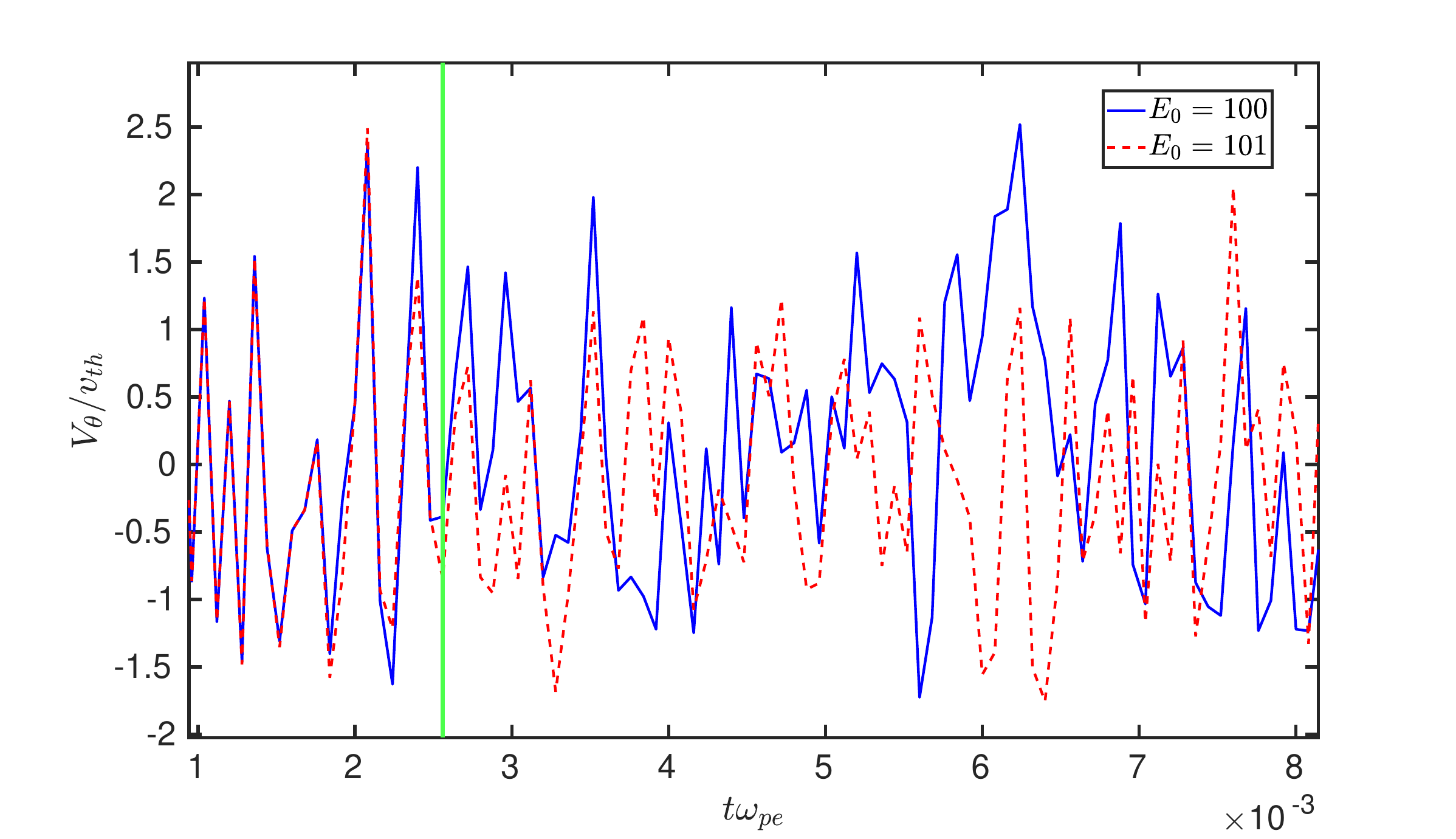}
     \caption{Time evolution of $V_\theta$ for one of the particles from the cluster for two different amplitude of the oscillatory electric field. The blue solid curve and red dashed curve correspond to the amplitude $100$ V/m and $101$ V/m, respectively.}
  \label{Fig:V_theta}
 \end{figure*}
  In Fig.\ref{Fig:V_theta}, the time evolution of $V_\theta$ has been shown for the two cases of slightly different amplitudes of the applied electric field. Here, the blue solid curve and red dashed curve correspond to two different amplitudes of oscillatory uniform electric field. Both the red and blue curves (corresponding to $E_0 = 101$ and $100$ respectively) remain similar till $0.00256$ $t\omega_{pe}$. This time is depicted by a green vertical line in the figure. Thereafter, the two curves become uncorrelated and display very different forms. In Fig.\ref{Fig:V_theta} the time intervals at which $V_{\theta}$ changes its sign, signifies the rotation reversal, which occurs at random intervals. Thus the rotational dynamics of the cluster is extremely sensitive to the amplitude of the applied external field. 
 
 We now analyze whether this random evolution has a chaotic characteristic or not. For this purpose, the techniques of time series analysis have been adopted. A time-delay coordinate ($\tau'$) for our time series $V_\theta$ which essentially measures the time duration after which the correlations in the signal die off.  The velocity autocorrelation function determines this.  This time delay coordinate helps in the reconstruction of the phase space attractor as shown by  Takens \cite{takens2006detecting} et. al. We determine the correlation dimension \cite{grassberger1983characterization} of the signal for different embedding dimensions. 
 \begin{figure}[hbt!]
   \includegraphics[scale=0.35]{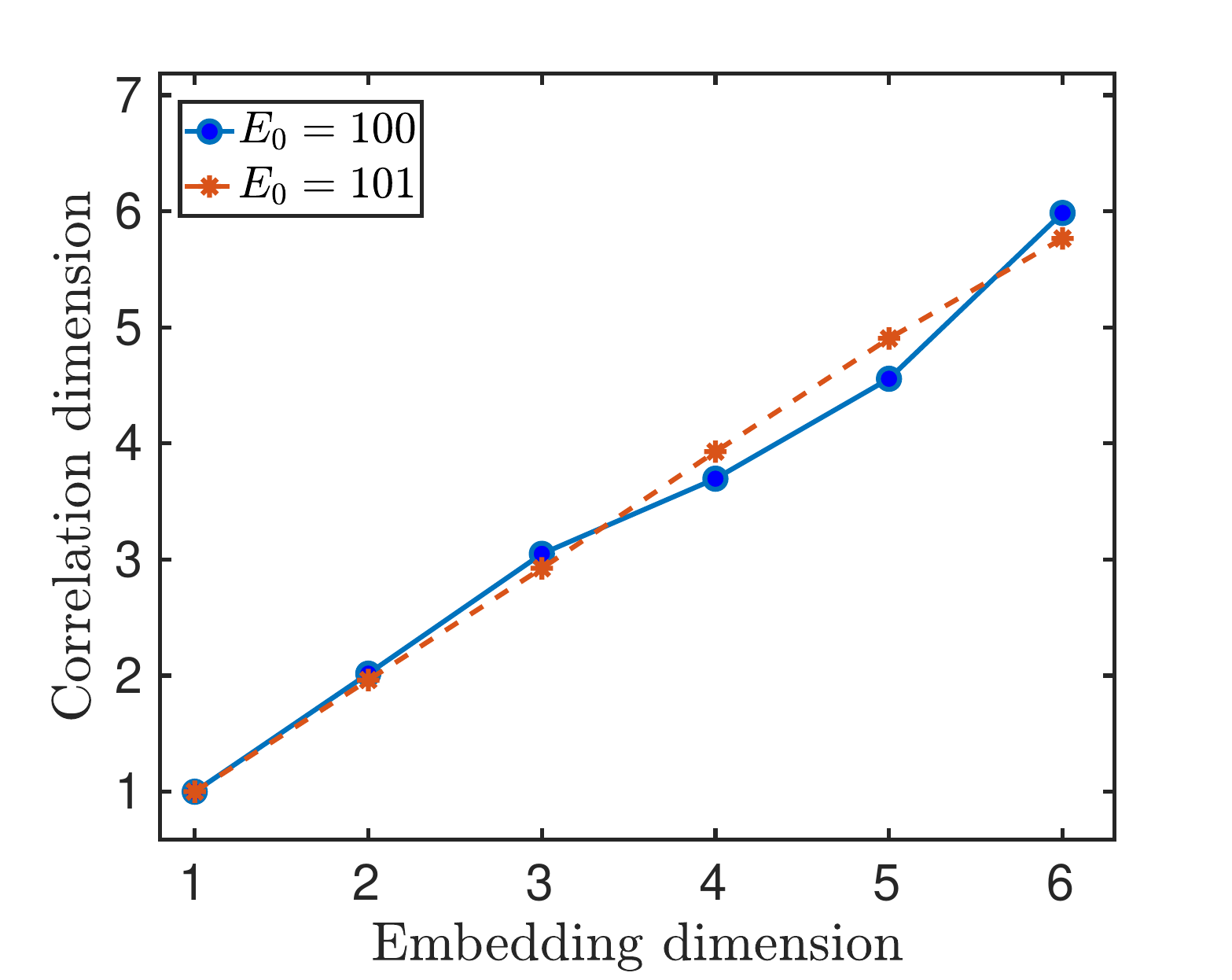}
    \caption{ Variation of correlation dimension with the embedding dimension for two amplitudes of the uniform electric field. The blue and red curves correspond to the $E_0$ $100$ V/m and $101$ V/m, respectively .}
  \label{Fig:correlation}
\end{figure}
 As shown in Fig.\ref{Fig:correlation} the correlation dimension increases linearly with the embedding dimension. There is no sign of the correlation dimension acquiring an asymptotic value as the embedding dimension is increased.  For chaotic behavior, the signal should asymptote at a low correlation dimension. Thus there is no evidence for the signal to be chaotic. Instead, the increasing value of the correlation dimension with the embedding dimension \cite{sheridan2005chaos} is indicative of the process to be stochastic (random).

\subsection{Parametric Studies} 
\begin{figure}[hbt!]
   \includegraphics[scale=0.33]{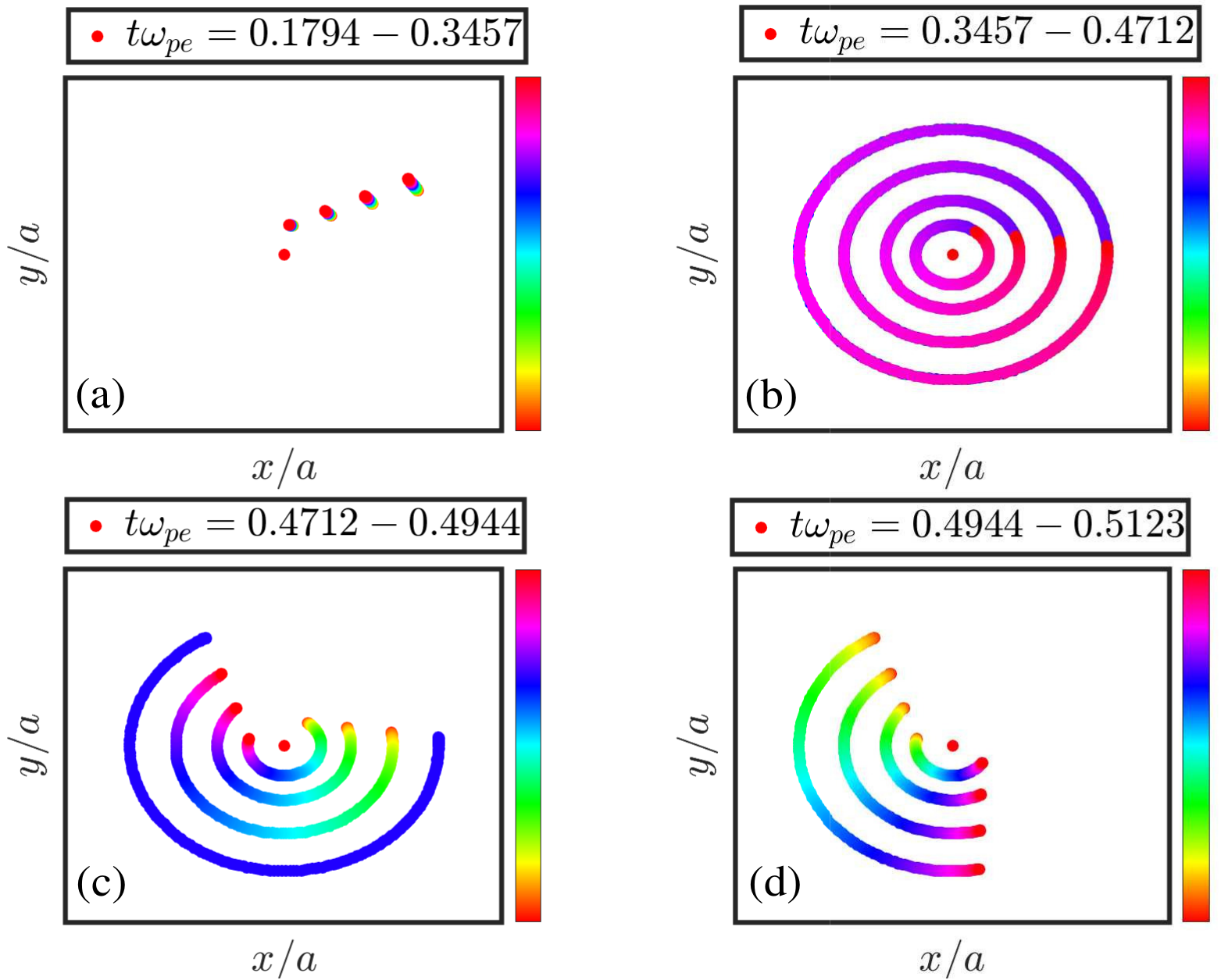}
     \caption{ Particle trajectory of one particle from each shell over the time duration $t\omega_{pe}= 0.1794-0.5123$. Different colors from orange to red show the time evolution of particles. The amplitude and frequency of the applied oscillatory uniform electric field are $E_0$ is $500$ V/m and $10\omega_{pe}$. } 
  \label{Fig:particle_trajectory500}
\end{figure}

\begin{figure}[hbt!]
   \includegraphics[scale=0.12]{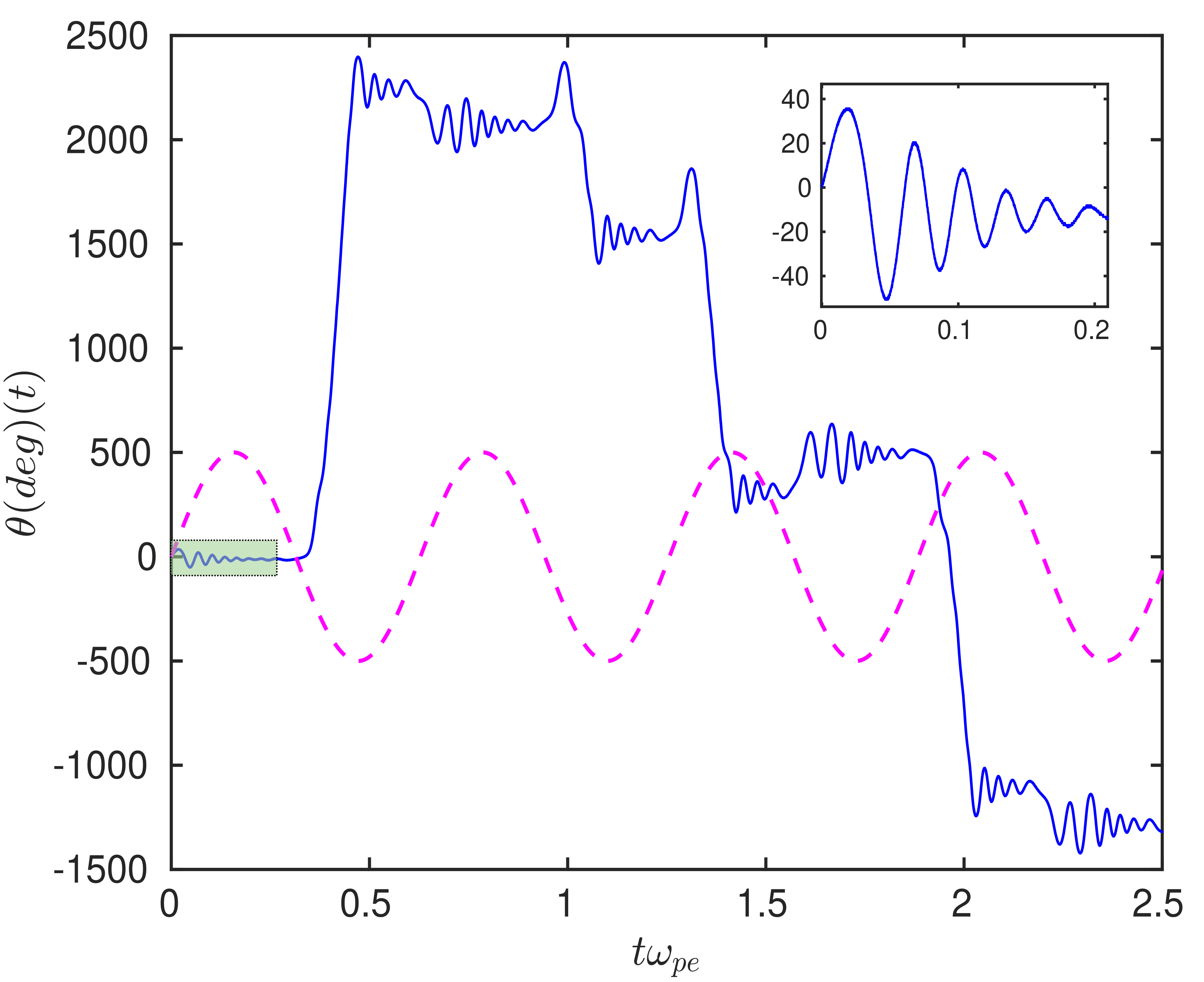}
     \caption{ The time evolution of angular displacement for one of the particles from the cluster having an amplitude and frequency of the oscillatory uniform electric field is $E_0$ is $500$ V/m and $10\omega_{pe}$. The inset represents the zoomed view of oscillations which is shown by colored box. The magenta curve shows the variation of the oscillatory uniform electric field ($E_0 sin(\omega t)$).} 
  \label{Fig:amplitude500}
 \end{figure}

The random behavior of the angular velocity persists even when the forcing amplitude is varied. When the amplitude of the force is increased the rotational velocity is faster and often covers several complete circulations before the reversal of the direction. There are also durations where the structure displays small oscillations. This can be seen from the plots in Fig.\ref{Fig:particle_trajectory500} and Fig.\ref{Fig:amplitude500} which show the particle trajectories in various shells and the evolution of the angle $\theta(t)$ respectively. 
\begin{figure}[hbt!]
   \includegraphics[scale=0.43]{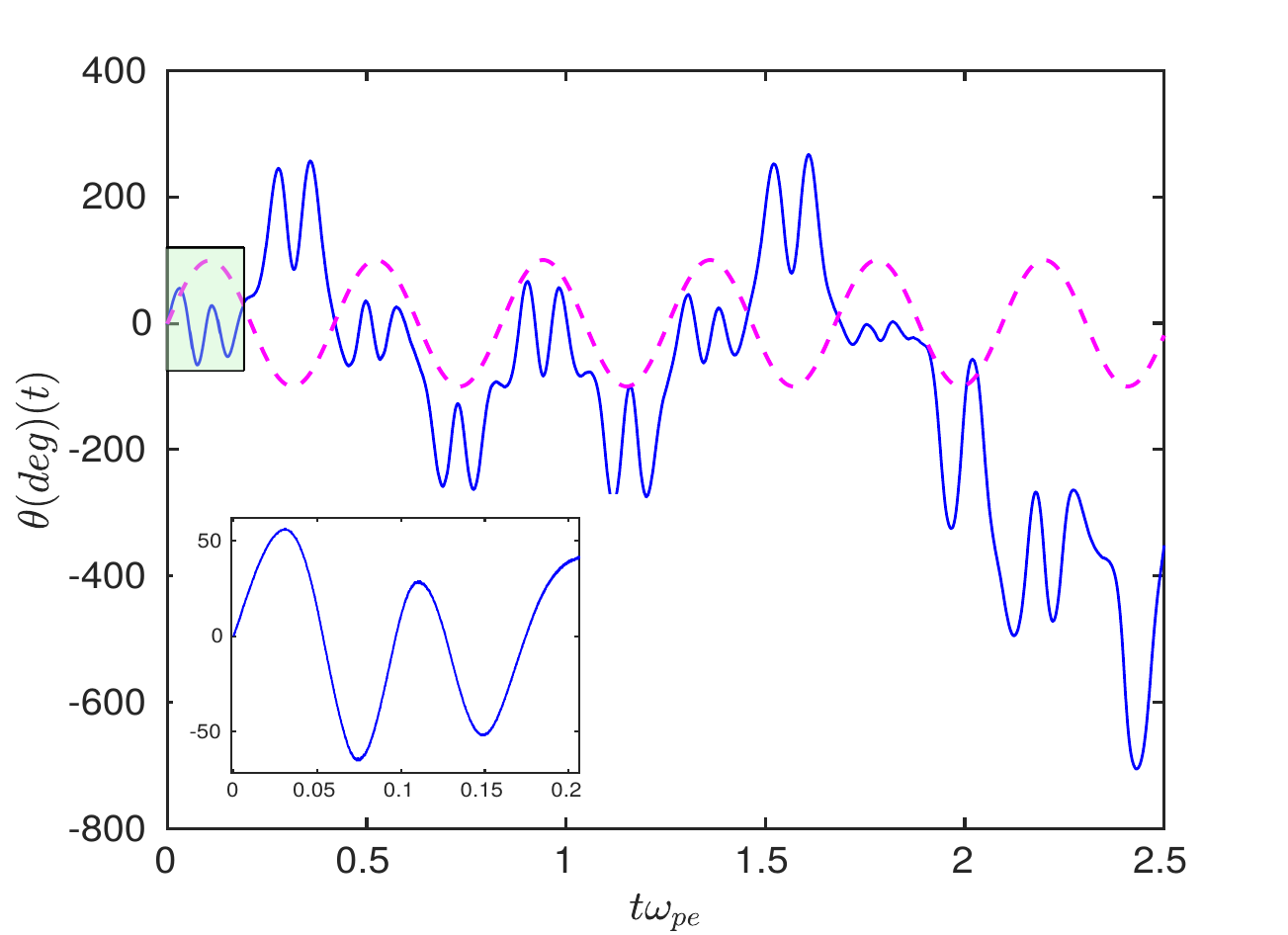}
     \caption{ The time evolution of angular displacement for one of the particles from the cluster having an amplitude and frequency of the oscillatory uniform electric field is $E_0$ is $100$ V/m and $15\omega_{pe}$. The inset represents the zoomed view of oscillations which is shown by colored box. The magenta curve shows the variation of the oscillatory uniform electric field ($E_0 sin(\omega t)$).} 
  \label{Fig:15wpe}
 \end{figure}

\begin{figure}[hbt!]
   \includegraphics[scale=0.65]{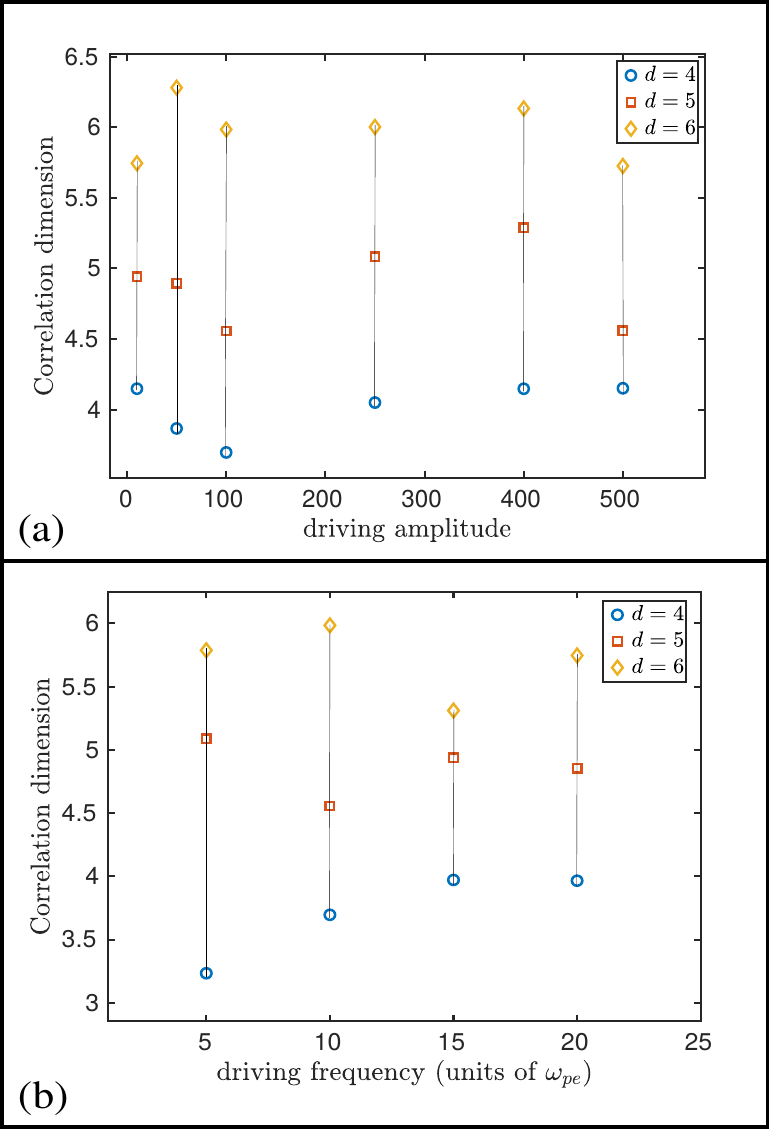}
    \caption{ (a) Variation of correlation dimension with the amplitude ($E_0$) of oscillatory electric field for three different embedding dimensions. (b) Variation of correlation dimension with the frequency ($\omega_0$) of the external applied oscillatory electric field. The circle, square, and diamond shapes correspond to four, five, and six embedding dimensions, respectively.  }
  \label{Fig:correlation_subplot}
\end{figure}

The characteristic behavior remains the same when the driving frequency is altered. The case of  ($\omega_0 = 15\omega_{pe}$) is shown in Fig.\ref{Fig:15wpe}. The time series analysis has also been carried out for various cases of driving amplitude and frequencies. It can be observed from Fig.\ref{Fig:correlation_subplot}, Subplot (a) and (b) show the variation of correlation dimension with driving amplitude and driven frequency (frequency in units of $\omega_{pe}$), respectively. The circle, square, and diamond shapes correspond to different embedding dimensions, e.g. four, five, and six respectively. It is clear from both the subplots that the correlation dimension increases with the embedding dimensions. 
     
 In the next section, we study the shielding in response to the time-varying charge on the 
 inserted particle. This can be viewed as a biased probe inserted in the medium with the bias voltage varying with time.

\section{Oscillatory charge as an external perturbation}
In this section, we study the behavior of crystalline structure when we introduce a normalized (with respect to electronic charge) oscillatory charge ($q(t) = q_0+q_1 sin(\omega_0 t)$)in the simulation box as an external perturbation. This external charge can be viewed as a Langmuir probe with time-varying biased potential. We choose the frequency   $\omega_0=10\omega_{pe}$ for all the cases of study here. Thus the forcing is chosen to be at a very high frequency compared to the natural response frequency of the plasma. The bulk plasma, therefore, does not get sufficient time to respond. We would see, therefore, that the crystalline shielding pattern evolves on its own. Often throwing particles out when the central charge reduces but never acquiring charge from the bulk region in the phase when the central charge increases.  

The value of central charge in steady state is chosen as $q_0 = 100$. Studies for various values of  $q_1$ have been performed. In particular, we will discuss the specific cases of   $q_1 = 20$ and $ 50$ here. The highest charge acquired by the central particle is thus $120$  and $150$ and the smallest value is $80 $ and $50$ for the choice of $q_1 = 20 $ and $50$ respectively. We start the time-dependent study for oscillating central charge once the shielding is complete for the steady case.  As discussed earlier the choice of high frequency ensures that the bulk plasma medium remains undisturbed. 

The evolution of the shielding cloud for the two cases of $q_1 = 20$ and $50$ can be observed in the attached movies as supplementary material.  We have attached the movie "Hexagonal.avi" \cite{movie2} as supplementary material to understand the dynamics of crystalline structure when ($q_1=20$). And we also prepared a movie "Pentagonal.avi" \cite{movie3} as a supplementary material which shows the transformation of crystalline structure from hexagonal to pentagonal structure when $q1=50$. 
For both cases in the phase when the {$q$} increases the cluster is pulled inwards and it starts rotating. The fast evolution of $q$ does not allow the accumulation of additional charges from the bulk plasma medium. During the phase when $q$ decreases the charges in the shielding cloud expand. During a few oscillations in the beginning some of the charge gets expelled to the bulk medium during this particular phase. Ultimately the cluster retains charges that are equal to the minimum value of $q$ for both cases. 

\begin{figure}[hbt!]
   \includegraphics[width=0.5\textwidth]{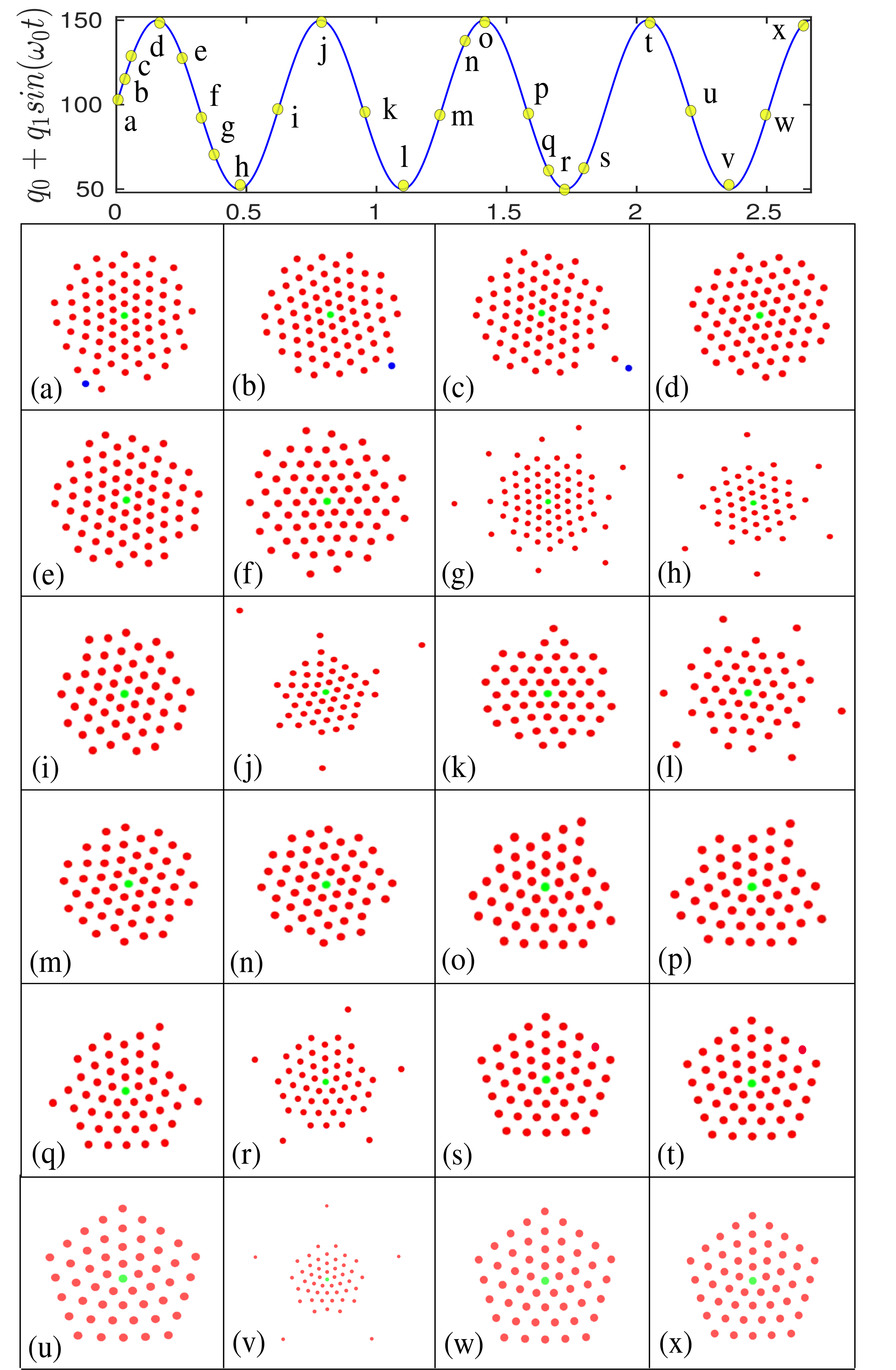}
     \caption{Schematic representation of the dynamics of the cluster over the time duration $t\omega_{pe}=2$. The blue sinusoidal curve shows the time variation of $q_0+q_1sin(\omega_0t)$ and the circle on this curve in yellow color represents the corresponding subplots. }
  \label{Fig:q50_dyno}
 \end{figure}

In Fig.\ref{Fig:q50_dyno} we show snapshots of the shielding arrangements at various times corresponding to the various values of $q$ depicted at the top by various points which have been depicted by the subplot alphabets. The $\hat {x}$ and $\hat {y}$ axes in the subplots are not identical and have been accommodated to suit the dimension of the cluster as it expands or contracts. The movie "Pentagonal.avi" \cite{movie3} in the supplementary material provides a better perspective in this regard. However, a couple of things are pretty clear from these subplots. In the first few cycles whenever  $q$ decreases the charges are expelled. Finally, the number of charges associated with the cluster acquires the same value as the minimum of $q$. At this phase, the crystal structure undergoes a transition and forms a pentagonal lattice structure (see subplot(s), (t) and (u)). For these subplots, $q$ the central charge is quite high compared to  $50$. The pentagonal lattice seems quite robust and stabilized. Even when $q$ reduces towards $50$ the lattice structure remains preserved, except that the particles in the periphery expand outwards (see subplot (v)). Thereafter, the structure again regains the compact pentagonal structure. 
We would like to point out here that for the case of $q_1 = 20$, the minimum charge $q = 80$ we do not observe the pentagonal structure formation. The $80$ remnant particles continue to be organized in hexagonal form.

 \begin{figure}[hbt!]
   \includegraphics[width=0.53\textwidth]{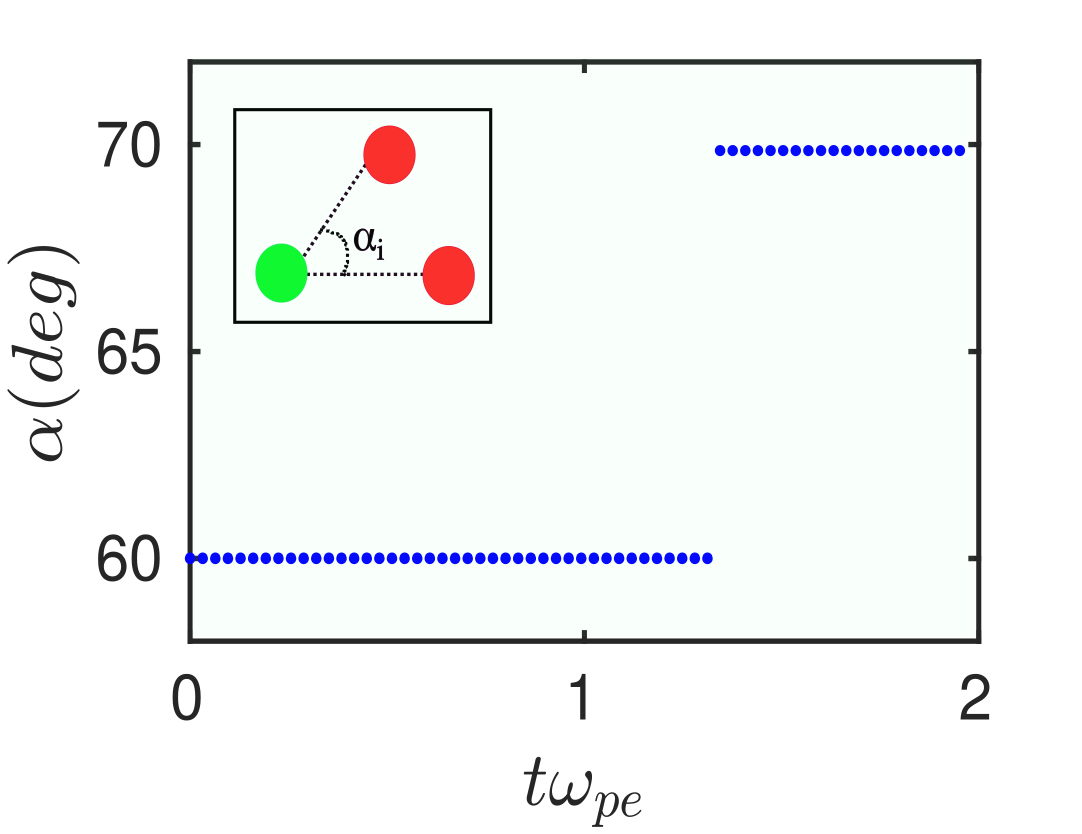}
     \caption{The variation of angle $\alpha$ (degree) values between two nearest neighbors (red dots) from a reference particle (green dot) along with normalized time. The inset shows how the value of $\alpha$ is calculated. The green and red dots represent the reference particle and the electrons near this reference particle, respectively.}
  \label{Fig:alpha}
 \end{figure}

The transition from hexagonal form to pentagonal structure can be quantified and also timed accurately by evaluating the mean angle $\alpha$ between nearest neighbors. 
\begin{equation}
    \alpha (deg) = \frac{1}{N_r}\sum_{i=1}^{N_r} \alpha_i 
    \label{equation:alpha_eqn}
\end{equation}
Where $N_r$ is the number of nearest particles around the reference particle, the $\alpha_i$ angle represents the angle between successive bonds formed by reference particles and their closest neighbors. The inset of Fig.\ref{Fig:alpha} shows how the value of $\alpha_i$ in degree is calculated. The green and red dots represent the reference particle and its nearest neighbors, respectively. From this, we can see that the initial angle remains at $60^0$ till $t\omega_{pe}=1.31$. This implies that the hexagonal structure prevails till this time. However, there appears a sudden transition in the value of  $\alpha$ to about $70^{o}$ which is within $2.7\%$ of $72^{o}$ interior angle of the pentagon, implying a transition to a pentagonal form.  We observe that when there is significantly less number of particles in comparison to the maximum charge $q$ remaining in the cluster due to a choice of very high $q_1$ the transition from hexagonal to pentagonal form occurs.


 \section{summary}
 We have investigated the dynamics of the shielding cloud around a high-charged static particle for a two-dimensional ultracold plasma in the presence of time-dependent fields ($\vec{E}=\hat{x} {E_0} sin(\omega_0 t)$). The molecular dynamics (MD) simulations are carried out for this purpose. Interesting rotational dynamics are observed which have been identified as having stochastic character. We have also applied external oscillatory perturbations in charge of the form ($q(t) = q_0+q_1sin(\omega_0t)$) on the particle around which the shielding cloud has gathered. A very interesting observation is the  
the transition from the hexagonal to the pentagonal lattice structure of the shielding cloud is observed when the value of $q_1$ is large.  
 
 It is thus clear that the ultracold plasmas display fascinating dynamics and characteristics.  It will be interesting to study and compare this with the behavior of one component of a strongly coupled dusty plasma medium which we are currently pursuing.  
 
 \begin{center}
    \textbf{ACKNOWLEDGEMENTS} 
 \end{center}
This research work has been supported by the Core Research Grant (Grant No. CRG/2022/002782) of the Department of Science and Technology, Ministry of Science and Technology, India. We also acknowledge support from the J.C. Bose Fellowship grant (Grant No. JCB-000055/2017)
from the Science and Engineering Board (SERB), Govt. of India. The authors thank the IIT Delhi HPC facility for computational resources. M.Y. is thankful to the University Grants Commission (Grant No. 1316/CSIR-UGC NET DEC.2017) for funding the research.


\bibliography{dynamics_ref}
\end{document}